\documentstyle[preprint,aps]{revtex}

\def\be{\begin{equation}}
\def\ee{\end{equation}}
\def\ci{\cite}
\def\bi{\bibitem}

\begin{document}

\preprint{}
\draft

\title{$Q^2$-dependence of semi-inclusive electron-nucleus
scattering and nucleon-nucleon correlations}

\author{O. Benhar $^a$, S. Fantoni $^b$, G.I. Lykasov $^c$
 and N.V. Slavin$^c$}
\address{$^a$ INFN, Sezione Sanita', I-00161, Roma, Italy \\
$^b$ Interdisciplinary Laboratory, SISSA,\\
and INFN, Sezione di Trieste, I-34014, Trieste, Italy \\
$^c$ Joint Institute for Nuclear Research,\\
141980 Dubna, Moscow Region, Russia}

\date{\today}

\maketitle

\begin{abstract}
We analize semi-inclusive electron-nucleus processes
$e + A \rightarrow e^{\prime} + h + X$ at moderate 
$Q^2$ and energy transfer $\nu$. Our results show 
that nucleons bound in the nuclear medium are 
distributed according to a function $f_A$ that reduces to the 
standard light-cone distribution in the Bjorken limit 
and exhibits a sizeable $Q^2$-dependence at lower $Q^2$, particularly 
when $Q^2 \sim \nu^2$. This distribution function is employed to 
calculate the cross section of the reaction 
$e + A \rightarrow e^{\prime} + p + X$, in the kinematical
region in which the detected proton is emitted backward
in the target rest frame.
\end{abstract}


\narrowtext

The semi-inclusive electron-nucleus scattering process 
$e + A \rightarrow e' + h + X$, in the kinematical region in which
the detected hadron $h$ is emitted approximatively backward 
in the target rest frame, has been recently investigated both 
experimentally \ci{deg} and theoretically \ci{bfls1}.
These studies were mainly aimed at pinning down the role of the 
high-momentum tails of the nucleon momentum distribution in nuclear
matter. In ref.\ci{bfls1} two approaches, based on different descriptions 
of nuclear structure at short interparticle distance, were compared. 
The first one was based on nuclear many-body theory, in which
the nuclear target is described in terms of nonrelativistic structureless 
nucleons whose 
distributon in momentum and removal energy is dictated by the nuclear spectral 
function $P({\bf k},E)$. Within the alternative approach, on the other hand, 
the relativistic invariant phase space volume availible to a quark
in a nucleus was computed taking into account nucleon-nucleon
correlations and the true Regge asymptotic of the quark distribution
in the nucleon.

In the first approach discussed in ref.\ci{bfls1} the spectral function 
is used to generate the distribution of nucleons in the nuclear 
target, $f_A$, as a function of the relativistic invariant quantity 
\be
z=\frac{(kq)}{(P_Aq)}\frac{M_A}{m}\ . 
\ee
In the above equation $P_A \equiv (M_A, {\bf 0}),M_A$ being the target mass,  
and $k \equiv (k_0,{\bf k})$ denote the initial four-momenta of the
target nucleus and the struck nucleon in the laboratory frame, 
respectively, $q \equiv(\nu,{\bf q})$ is the four-momentum transfer
and $m$ is the nucleon mass. The regime of deep inelastic scattering
(DIS) is defined by the Bjorken limit $Q^2, \nu \rightarrow \infty$ 
with $x=Q^2/2m\nu$ finite ($Q^2 = |{\bf q}|^2 - \nu^2 = -q^2$). In this 
limit the variable $z$ reduces to the ordinary light-cone variable: 
\be
z \rightarrow z_{Bj} = \frac{k^{+}}{m} = \frac{(k_0 - k_z)}{m}.
\ee
However, it has been recently pointed out \ci{bfls2} that 
when the value of $Q^2$ is not large enough and the 
Bjorken limit is not yet reached the quantity $z$ cannot be identified
with the light cone variable. As a consequence, the
distribution function $f_A$ depends upon both $z$ and 
$\beta~=~(|{\bf q}|/\nu)=(1~+~4 m^2 x^2/Q^2)^{1/2}$. The results
of ref.\ci{bfls2} show that this $\beta$-dependence is sizeable, 
particularly at $z > 1$. 

The semi-inclusive spectra of $\pi$ mesons 
and protons discussed in ref.\ci{deg} have been produced scattering 
14.5 GeV electrons and positrons off CO at moderate $Q^2$ and $\nu$. 
The kinematical domain covered by the experiment corresponds to 
energy loss $0.2~< \nu~<~12.5$~GeV and $Q^2$ in the range 
$0.1 < Q^2 < 5.0 (GeV/c)^2$. Hence, the beta dependence of the distribution
function $f_A$ is expected to be relevant to the theoretical analysis of the
data of ref.\ci{deg} and has to be quantitatively taken into account.
In this note we extend the calculations of ref.\ci{bfls1}, including the
dependence of $f_A(z,\beta)$ upon $\beta$ in the analysis of 
semi-inclusive $e + A \rightarrow e' + h + X$ processes at moderate $Q^2$ 
and $\nu$. 

The distribution function $f_A(z,\beta)$ is defined as:
\be
f_A(z,\beta)= z \int d^4k\ S(k)\ \delta\left(z-\frac{(kq)}{(P_Aq)}
\frac{M_A}{m}\right)\ ,
\label{1}
\ee
where $S(k)$ denotes the relativistic vertex function, that can be 
approximated
by the nonrelativistic spectral function $P({\bf k},E)$ according to \ci{bps}
\be 
S(k)=\left(\frac{m}{k_0}\right)P({\bf k},E)\ ,
\label{2}
\ee
with
\be
k_0=M_A - [(M_A - m + E)^2 + \mid {\bf k}^2\mid]^{1/2}.
\label{3}
\ee
The details of the calculation of $f_A(z,\beta)$ are presented 
elsewhere \ci{bfls2} and will be omitted here for the sake of 
conciseness. The numerical results of ref.\ci{bfls2}, corresponding to
infinite nuclear matter at equilibrium density, are summarized in 
Fig. 1, where  the $z$ behaviour of the function $f_A(z,\beta)$ is shown for 
different values of $\beta$. It clearly appears that $f_A(z,\beta)$ strongly
depends upon $\beta$ at $z > 1$.

Let us now turn to the semi-inclusive momentum spectrum of the protons 
emitted in the $e + A \rightarrow e^\prime +  p + X$ around the backward 
direction, in the kinematical region forbidden to the quasi-free 
electron-nucleon process. At small transerse momenta, i.e. at  
$p_t \sim 0$, this spectrum, that will be denoted  $\rho(z,x,Q^2)$,  
is proportional to the distribution of the nucleons in the target as a 
function of the variable $z$, defined as in eq.(1). According to the two
approaches discussed in ref.\ci{bfls1} this distribution can be identified
with either the function $f_A(z,\beta)$ of eq.(3) or with a function 
$T_A(z)$, constructed under the assumption that the nucleus can be viewed 
as a collection of multiquark bags whose structure functions exhibit 
Regge asymptotic behaviour \ci{Efr}. Within the framework of the latter 
approach the hadron yields measured in hadron-nucleus collisions in the 
kinematical region forbidden to hadron-nucleon inelastic scattering 
, the so-called limiting fragmentation processes, can be 
quantitatively accounted for \ci{Efr,Lyk,Efr2}. However, in this picture
the argument of the distribution function $T_A$ is the usual light-cone
variable $z_{Bj}$, that coincides with $z$ of eq.(1) in the Bjorken limit 
only. One has:
\be
\rho(z,x,Q^2) \equiv \int d^2p_t\ E_p\  
\frac{d\sigma(e + A \rightarrow e^{\prime} + p + X)}
{d^3p\ dE_{e^\prime}\ d\Omega_{e^\prime}} \sim T_A(z_{Bj}) F^N_2(x,Q^2)\ ,
\label{rho_2}
\ee
where $F^N_2(x,Q^2)$ is the nucleon structure function.
On the other side, the first approach discussed in ref.
\ci{bfls1}, based on the use of the nuclear spectral function, leads
to the following expression:
\be
\rho(z,x,Q^2) \sim f_A(z,Q^2) F^N_2(x,Q^2)\ ,
\label{rho_1}
\ee
with $f_A$ defined as in eq.(3) (notice however that in the above equation, 
the dependence upon $\beta$ has been replaced with an explicit dependence 
upon $Q^2$).

In order to compare the spectra defined by eqs.(\ref{rho_2}) and (\ref{rho_1}) 
to the data of ref.\ci{deg} one has to integrate over the relevant ranges of
$Q^2$ and $\nu$, as done in ref.\ci{bfls1}. The integrated spectra are shown 
in Fig. 2 as a function of the kinetic energy of the detected proton, $T$, 
emitted at angle $\Theta = \arccos(-0.55)$ with respect to the direction of the
thre-momentum transfer ${\bf q}$. The calculations have been performed using 
the empirical nucleon structure function $F_2^N(x,Q^2)$ at $0.2~<~Q^2~<~5$
(GeV/c)$^2$, extracted from electron-proton and 
electron-deuteron data according to the parametrization of ref.\ci{Cap}.
 
The solid line in Fig. 2 represents the integrated proton spectrum 
computed using eq.(\ref{rho_1}), with $f_A(z,\beta)$ given by eqs.(3)-(5), 
and the nuclear matter spectral function of ref.\ci{BFF}. The calculation
 has been 
carried out for proton kinetic energies $T \leq T_{max}$, with 
$T_{max}\sim$~0.28~ GeV, that corresponds to $z\leq1.9$. This cutoff reflects
the momentum range $k \leq $~0.8~(GeV/c) covered by the spectral function of 
ref.\ci{BFF}. The effect of the dependence of the distribution function
upon $\beta$ (or, equivalently, upon $Q^2$) can be estimated comparing the solid
line to the dashed line, which has been obtained setting $\beta=1$, i.e. 
identifying $z$ of eq.(1) with the light-cone fraction of eq.(2).
The two curves significantly differ, indicating that, as expected, in 
 the moderate $Q^2$ range of the data of ref.[1] the Bjorken limit is
not applicable and the $Q^2$ dependence associated with the distribution
function cannot be disregarded. 

In Fig. 2 we also show, for comparison, the integrated proton spectrum 
computed from eq.(\ref{rho_2}) with the 
distribution function $T_A$ discussed in ref.\ci{bfls1}.  
Comparison between the solid and dot-dash lines suggests 
 the existence of a connection between
the processes $h + A \rightarrow p + X$ and 
$e + A \rightarrow e^{\prime} + p + X$ when the produced 
 proton is emitted backward, i.e. in the kinematical domain
forbidden to hadron-nucleon and electron-nucleon collisions.

In conclusion, the results of our work, summarized in Figs. 1 and 2, 
show significant dependence of the nucleon distribution $f_A$ 
upon the magnitude of the virtual photon velocity
$\beta=|{\bf q}|/\nu$, or equivalently upon $Q^2$, at fixed $x$. The 
inclusion of this effect in the theoretical analysis of 
semi-inclusive backward proton production in the kinematical regime of the
experiment reported in ref.[1] leads to an improvement of the agreement
between theory and data up to $T\sim$ 2.8 GeV, corresponding to the
maximum quantity $z(Q^2)\sim$1.9. 
Unfortunately, the limitations inherent to the nonrelativistic nature of
the many-body theory employed in the calculation of the spectral function 
make it impossible to extend the calculation of $f_A(z,\beta)$ to larger 
values of $z$.

The results obtained from eq.(\ref{rho_2}) using the nucleon 
distribution $T_A$ dictated by true Regge
asymtotic at $z_{Bj} \rightarrow 2$ and $z_{Bj} \rightarrow 3$ \ci{Efr,Lyk} 
are also in satisfactory agreement with the data at $T<$ 2.8 GeV. In addition, 
they account for the qualitative behaviour of the experimental spectrum 
up to $T \sim $ 0.4 GeV, corresponding to $z_{Bj} \sim$ 2.5.
A detailed theoretical analysis of the relationship between the functions 
$f_A$ and $T_A$ may therefore provide some guidance on the
the extrapolation of the nuclear spectral function to 
the large $k$ domain, inaccesible within nonrelativistic nuclear many-body
theory.
 
\acknowledgments

This work had been encouraged and supported by the Russian Foundation
of Fundamental Research. We gratefully acknowledge very helpful
discussions with A. Fabrocini. One of us (GIL) wishes to thank 
S. Fantoni for the kind hospitality at the Interdisciplinary
Laboratory of SISSA, where part of this work has been carried out.

\begin{figure}
\caption{
$z$-dependence of the distribution function $f_A(z,\beta)$ 
of eq.(3) at different values of $\beta$. Solid line: $\beta=1$; 
dashed line: $\beta=1.1$; dot-dash line: $\beta=1.2$; dotted 
line: $\beta=1.3$.}
\end{figure}

\begin{figure}
\caption{
Kinetic energy spectra of protons emitted at angle $\Theta$ in the
semi-inclusive recation $e + CO \rightarrow e^\prime + p + X$. 
The solid curve corresponds to the full calcutaion
carried out using eq.(7), whereas the dashed line represents the
results obtained setting $\beta$=1. The dot-dash line shows the
spectrum calculated from eq.(6) using the nucleon distribution $T_A$
of ref.[5]. The experimental data are taken from ref.[1].}
\end{figure}


\begin{references}

\bi{deg}
P.V. Degtyarenko {\it et al}, Phys.Rev C, {\bf 50}, R541 (1994).
\bi{bfls1}
O. Benhar, S. Fantoni, G.I. Lykasov and N.V. Slavin,
Phys. Rev. C, {\bf 55}, 244 (1997).
\bi{bfls2}
O. Benhar, S. Fantoni, G.I. Lykasov and N.V. Slavin,
preprint SISSA-ISAS 78/97/CM (1997).
\bi{bps}
O. Benhar, V.R. Pandharipande and I. Sick, Phys. Lett. B, in press.
\bi{Efr}
A.V. Efremov, A.B. Kaidalov, V.T. Kim, G.I. Lykasov and N.V. Slavin, 
Sov. Journ. Nucl. Phys., {\bf 47}, 868 (1988).
\bi{Lyk}
G.I. Lykasov, Phys. Part. Nuclei, {\bf 24}, 59 (1993).
\bi{Efr2}
A.V. Efremov, A.B. Kaidalov, G.I. Lykasov and N.V. Slavin, Sov. Journ. 
Nucl. Phys., {\bf 57}, 874 (1994).
\bi{Cap}
A. Capella, A. Kaidalov, C. Merion and J. Tran Than Van, 
Phys. Lett., {\bf B337}, 358 (1994); {\bf B343}, 403 (1995).
\bi{BFF}
O. Benhar, A. Fabrocini and S. Fantoni, Nucl. Phys., {\bf A505}, 267 (1989).
\end{references}
\end{document}